\documentclass[sigconf]{acmart}

\usepackage{tabularx}
\usepackage{enumitem}
\usepackage[shortcuts]{extdash}

\AtBeginDocument{%
  \providecommand\BibTeX{{%
    \normalfont B\kern-0.5em{\scshape i\kern-0.25em b}\kern-0.8em\TeX}}}

\copyrightyear{2024}
\acmYear{2024}
\setcopyright{acmlicensed}\acmConference[ExEn '24 ]{2024 Workshop on Explainability Engineering}{April 20, 2024}{Lisbon, Portugal}
\acmBooktitle{2024 Workshop on Explainability Engineering (ExEn '24 ), April 20, 2024, Lisbon, Portugal}
\acmDOI{10.1145/3648505.3648507}
\acmISBN{979-8-4007-0596-0/24/04}

\begin{document}

\title{Generating Context-Aware Contrastive Explanations in Rule-based Systems}

\author{Lars Herbold}
\orcid{0009-0003-1292-1528}
\affiliation{%
  \institution{University of Cologne}
  \city{Berlin}
  \country{Germany}
}
\email{lars.herbold@outlook.de}

\author{Mersedeh Sadeghi}
\orcid{0000-0001-6405-8824}
\affiliation{%
  \institution{University of Cologne}
  \city{Cologne}
  \country{Germany}}
\email{sadeghi@cs.uni-koeln.de}

\author{Andreas Vogelsang}
\orcid{0000-0003-1041-0815}
\affiliation{%
  \institution{University of Cologne}
  \city{Cologne}
  \country{Germany}}
\email{vogelsang@cs.uni-koeln.de}

\renewcommand{\shortauthors}{Herbold et al.}

\begin{abstract}
Human explanations are often contrastive, meaning that they do not answer the indeterminate ``Why?'' question, but instead ``Why~P, rather than Q?''. Automatically generating contrastive explanations is challenging because the contrastive event (Q) represents the expectation of a user in contrast to what happened. We present an approach that predicts a potential contrastive event in situations where a user asks for an explanation in the context of rule-based systems. Our approach analyzes a situation that needs to be explained and then selects the most likely rule a user may have expected instead of what the user has observed. This contrastive event is then used to create a contrastive explanation that is presented to the user. We have implemented the approach as a plugin for a home automation system and demonstrate its feasibility in four test scenarios.     
\end{abstract}



\maketitle

\section{Introduction}
\label{sec:Intro}
As smart cyber-physical systems, complex decision-making, and autonomous systems have advanced rapidly in recent years, the need for explainability in these domains has also increased. Explainability is a crucial research area that aims to provide transparent and understandable reasons for the outcomes and behaviors of these systems~\cite{chazette2020explainability}. Users need to understand how the system works and why it behaves in certain ways, otherwise, they may lose confidence and trust in it. This can result in negative consequences such as misuse or rejection of the system~\cite{lim2009and}. One way to address this issue is to \textit{explain} the system's rationale and outcomes to the users. This approach has been proven to work well in previous studies: A common finding in the literature is that trust and interaction quality between humans and computers can be improved by providing explanations~\cite{winikoff2018towards}. 

Contrastive explanations clarify why an event occurred in contrast to another. They are intuitive for humans to produce and comprehend~\cite{lipton1990contrastive, jacovi2021contrastive}. There is evidence that explanations by humans are often contrastive~\cite{artelt2021contrastive, miller2021contrastive}. Contrastive explanations can provide a more accurate and fine-grained interpretability of system decisions. User-friendly explanations are important in smart environments because they can help users to better understand and control the technology, which can drive the adoption of smart technology~\cite{dai2023effect, reisinger2022user}.
Numerous studies support the view that an explanation is not a one-size-fits-all concept. Instead, an individual’s characteristics, expectations, prior knowledge, and other contextual factors play a role in determining what they consider to be a satisfactory explanation~\cite{ferreira2020people, liao2020questioning, miller2019explanation}. Smart environments can enhance user engagement and satisfaction, and ultimately achieve greater success, by offering explanations that are tailored to the individual and relevant to the context~\cite{ferreira2020people}.

Despite the theoretical and empirical studies that show the value of contrastive explanations~\cite{miller2019explanation, miller2021contrastive, reisinger2022user, winikoff2018towards}, there is a lack of approaches that show how to generate contrastive explanations for users. Some work has been done in the context of Explainable~AI~(XAI)~\cite{chen2021kace, stepin2021survey}. Explanation generation approaches for rule-based smart environments have, so far, focused on generating causal explanations (i.e., explaining the reasons for a happened event)~\cite{houze2022generic, nandi2016automatic}.
This paper proposes an approach for generating context-aware contrastive explanations in smart environments. Users generally need explanations after an event has occurred that does not align with their expectations~\cite{lipton1990contrastive}. Therefore, conservative explanations are inherently context- and user-specific. Our approach analyzes the situation in which an explanation is requested and predicts a possible expected event by a user. This contrastive event is then used to create a contrastive explanation that is presented to the user. We have implemented the approach as a plugin for a home automation system and demonstrate its feasibility in four test scenarios.

\section{Background}
\label{sec:Background}

\subsection{Explainability and its Engineering}
Explainability (i.e., the ability of a system to provide explanations for its behavior) has been framed as a key requirement for software-supported decisions and a means of promoting transparency~\cite{chazette2020explainability}. Explanations are seen as an option to mitigate the lack of transparency in a system. They can improve the understanding of a system by conveying information, thus influencing interpretability or understandability. Explainability engineering covers all activities related to making a system explainable~\cite{Brunotte2022}. This includes the elicitation and specification of explanation needs and related requirements~\cite{Unterbusch23,Sadeghi21}, the (automatic) detection of explanation needs at runtime, the construction of context-aware explanations, and a user-specific presentation.

\subsection{Contrastive Explanations}

Contrastive explanations are one type of explanation that explains why an event $P$ occurred rather than event $Q$~\cite{lewis1986causal}. 
In common models of contrastive explanations, such as those of Lipton~\cite{lipton1990contrastive} and Miller~\cite{miller2021contrastive}, there are two central concepts: 
\begin{itemize}[leftmargin=*]
    \item A \emph{fact} is a piece of information that is considered to be true and can be verified. In the context of scientific explanations, a fact refers to a specific aspect of an event or phenomenon that is being explained. It is not simply the event or phenomenon itself, but a particular characteristic or feature of it that is of interest.
    \item A \emph{foil} is an alternative scenario or situation that is used as a point of comparison to the fact. In the contrastive analysis of an explanation, the foil is used to focus the explanation by providing a further restriction on explanatory causes. The same fact may have several different foils, and the causes that explain a fact relative to one foil will not generally explain it relative to another. For example, when explaining why leaves turn yellow in November, we may not be interested in explaining this phenomenon in general, but only in comparison to a specific alternative scenario, such as why they turn yellow in November rather than in January or why they turn yellow rather than turning blue.
\end{itemize}

The \emph{fact} relates to a situation \emph{as is}. The \emph{foil} to an imaginary situation, that could have been. Lewis states that a contrastive explanation provides information for why the \emph{fact} occurred, but the \emph{foil} did not~\cite{lewis1986causal}.
Related to contrastive explanations are counterfactual explanations, which explain which past events would have caused the \emph{foil} to materialize~\cite{roese1997counterfactual}.

\subsection{Smart Environments and Rule-based Systems}
A \emph{smart environment} is a physical space that has been enhanced with technology, such as sensors, actuators, and computing devices, to gather and use information about the environment and its occupants to adapt to their needs and preferences~\cite{el2021smart}. Alam~et~al.~\cite{alam2012review} describe a smart environment's abilities to include assisting inhabitants to live independently and comfortably with the help of technology and providing context-aware automated or assistive services for the user. This usually includes automating tasks with the help of this technology~\cite{el2021smart, nandi2016automatic}.
In our work, we refer to smart homes and smart offices as smart environments although the term is used more broadly in other studies and can include applications such as smart farming, smart cities, smart education, or smart factories.
Smart environments can incorporate a set of \emph{rules} to automate and control various aspects of its physical space. Similar to the concept of Nandi and Ernst~\cite{nandi2016automatic}, a rule consists of two sets: \emph{Preconditons} and \emph{actions}. A \emph{precondition} is a binary indicator about the state of a part of the smart environment. Preconditions can be combined by logical operators to form more complex preconditions. When the precondition of a rule is fulfilled, the rule \emph{fires} and executes an \textit{action}. Systems that rely on a set of rules to execute their decisions are called \emph{rule-based systems}.

\section{Approach}

The main challenge for building a contrastive explanation is identifying the foil, which represents the user's expectation in contrast to what happened (the fact). The foil is usually user- and context-specific, i.e., for the same fact, the foil may differ depending on the user and the situation. In rule-based systems, facts and foils refer to actions that the system performed as a result of a rule being fired (or not fired in the case of a foil). Therefore, in our approach, a foil always refers to an action that was not performed by the system but that was expected by the user. The objective of our approach is to identify the action and the corresponding rule of a rule-based system that represents the foil for a user in a specific situation. The identification of the need for a (contrastive) explanation is not in the scope of our approach. That means we assume that our approach is triggered in appropriate situations (e.g., by a user requesting an explanation or the system detecting a need for an explanation~\cite{Blumreiter2019}). The identification of the fact is also not in the scope of our approach. We assume that the identification of the fact (the action that caused the explanation need) is straightforward as it may be the last action that happened or a specific action for which the user requests an explanation. 
In earlier work, we have provided a conceptual model to determine the \emph{fact} in smart environments~\cite{sadeghi2024}.

\begin{figure}
    \centering
    \includegraphics[width=\linewidth]{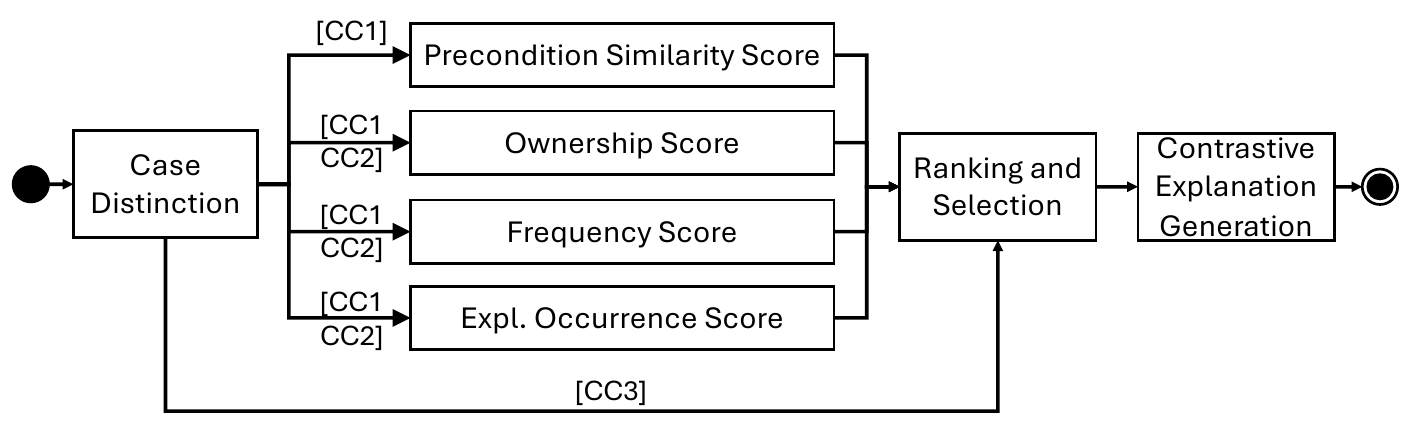}
    \caption{Overview of our approach for contrastive explanation generation. Depending on three possible confusing cases (CC), different scores are ranked to select the most likely expected alternative event.}
    \label{fig:framework}
\end{figure}

Figure~\ref{fig:framework} provides an overview of our approach. In the first step, we distinguish between three different confusing cases (CC) in which a contrastive explanation may be requested. These three different cases determine the set of factors that may influence the determination of the foil. Our approach then considers four factors that assign scores to all existing rules in the system. Each score contributes to the overall likelihood that a specific rule of the system was expected by a user in a certain situation. In the \textit{ranking and selection} step, the approach accumulated the scores and selects the \textit{most likely expected} rule. Based on this rule, of which the action represents the foil, an explanation can be generated that contrasts the fact (what happened) with the prospective foil (what was expected instead).
In the following, we describe the steps in detail.

\subsection{Case Distinction based on the \textit{Fact}}
\label{subsec:Framework/cases}
We identified three basic \emph{confusing cases (CC)} where the user may expect a different outcome. In each case, the user expects an event to happen as a result of a rule being fired. The \emph{fact}, in turn, is different in the three confusing cases.

\begin{description}[leftmargin=*]
    \item[\textbf{CC1:}] \emph{An action occurred but another action was expected}:
    The user is confused by an action of a fired rule that is different from the user's expectations. 
    In this case, the \emph{fact} is the action of the rule that was fired, and the \textit{foil} is the expected action.
    
    \item[\textbf{CC2:}] \emph{No action occurred but an action was expected}:
    The user is confused because the system did not perform an expected action. 
    This may be due to two reasons. Either the preconditions of the expected rule were not met and the rule was not fired (and no other rule regarding the device), or the \emph{expected rule} does not exist anymore in the system (e.g., it was deleted). 
    We will not go into detail about the case if a rule was deleted but outline later (Section~\ref{subsec:Framework/Discussion}) what could be done in future work. 
    In both cases, the \emph{fact} is ``nothing,'' and the \textit{foil} is the expected action.
    
    \item[\textbf{CC3:}] \emph{An error occurred but an action was expected}:
    An error occurred, so the resulting situation is different from the user's expectation. 
    In this case, the \emph{fact} is the occurring error, and the \textit{foil} is the expected action.
\end{description}

In each case, a phenomenon occurred that caused the underlying confusing situation. In general, we call this the \emph{happened event} and treat it as an equivalent to the \emph{fact} in our smart environment setting.

\subsection{Calculating \textit{Expected Rule} Scores}
\label{subsec:Framework/measures}
To determine the \textit{expected rule}, we define four factors that may influence the user's expectation:
\begin{enumerate}[leftmargin=*]
    \item \textbf{Precondition Similarity}: Actions of rules with similar preconditions may be more likely to be expected. 
    \item \textbf{Ownership}: Actions of rules that have been created by a user may be more likely to be expected.
    \item \textbf{Frequency}: Actions of rules that fire frequently may be more likely to be expected.
    \item \textbf{Explanation Occurrence}: Actions of rules that have been explained before may be more likely to be expected.
\end{enumerate}

In \emph{CC1}, all four factors can be applied to assess the \emph{candidate rules}. Since \emph{precondition similarity} can only be calculated if the \emph{fact} is a rule, it cannot be used in the other cases. For \emph{CC2}, the remaining three factors are applied. In case of an error (\emph{CC3}), we know that the confusion was caused by the appearance of the error. Thus, we determine the \emph{expected rule} as the last rule fired before the error occurred. Therefore, we do not consider any \textit{expected rule} score in this case.
The following paragraphs provide a detailed explanation of each factor and what value it contributes to a candidate rule. 

\textbf{Precondition Similarity.}
Actions that result from rules with similar preconditions to the rule that caused the observed action (the fact) are good candidates for a foil because a user may falsely assume that the preconditions of another rule are met and therefore, they expect the action of that other rule.
Precondition similarity can only applied in situation \emph{CC1}, where the \textit{fact} is the result of a rule that was fired. 
In our approach, we compare all rules of a system with the rule that created the fact and calculate a precondition similarity score.
The more similar the preconditions of two different rules are, the more likely it is that the user expects one of these rules if the other is fired. 

To determine the similarity between the precondition of the candidate rule and the happened rule $s(\mathit{CR},\mathit{HR})$, we examine the set of unique conditions in the precondition of the candidate rule $P_{CR}$ and the set of unique conditions in the precondition of the happened rule $P_{\mathit{HR}}$. We calculate the Jaccard similarity between these two sets, which is defined as:

$$\mathit{Jaccard}(P_{CR},P_{HR})=\frac{|P_{CR}\cap P_{HR}|}{|P_{CR}\cup P_{HR}|}$$

Consider an example with two rules: Rule $A$ (the candidate rule) turns a fan on if the temperature in the room is higher than 25\textcelsius~and someone is in the room. Rule $B$ (the happened rule) turns the air conditioning on if the temperature in the room is higher than 25\textcelsius, it is a summer month, and someone is in the room. Rule $A$ has two preconditions ($|P_A|=2$) and rule $B$ has three preconditions ($|P_B|=3)$. The cardinality of the union of preconditions of both rules is $3$ ($|P_A \cup P_B|$=3). Of these three preconditions, rule $A$ and rule $B$ share two ($|P_A \cap P_B|=2$). The precondition similarity of $A$ and $B$ is now calculated as follows:

$$s(A,B)=\mathit{Jaccard}(A,B)=\frac{|P_{A}\cap P_{B}|}{|P_{A}\cup P_{B}|}=\frac{2}{3}$$

\textbf{Ownership.}
If the \emph{explainee} created a rule in the system, it is more likely that they expect this rule to be fired. We call the creating user of a rule the \emph{owner} of the rule. Because of the ownership, we assume, they have a good understanding of the rule's functionality. Ownership is a binary measure with a value of $1$  if the user is the owner of the rule or $0$ if the user is not the owner of the rule.

\textbf{Frequency.}
The more often a rule is fired by a system, the more likely it is that the explainee has experienced the rule’s actions before. A higher frequency therefore makes a rule more expected by users than rules that are rarely fired and therefore have not been experienced as often by users. Based on the history of fired rules, we assign the number of times a rule was fired in a certain period as the frequency score.

\textbf{Explanation Occurrence.}
If the system has already explained a rule to the user before, the user is more aware of the existence and functionality of this rule. Thus, the user is more likely to expect this rule in comparison to other, less explained rules. Like the frequency score, we assign the number of times a rule has been explained in a certain period as the occurrence score.

\subsection{Deciding for an Expected Rule}
\label{subsec:Framework/rating}

Each rule of the system is now characterized by a set of values that indicate the expectation likelihood based on the set of up to four factors mentioned above. These values now need to be weighed and compared to each other to derive a ranked list of rules and finally decide on the most likely expected rule. This is a typical multi-criteria decision-making problem and we apply TOPSIS to rank the alternatives~\cite{hwang1993new, zulqarnain2020application}. TOPSIS is based on the concept that the chosen alternative should have the shortest geometric distance from the positive ideal solution and the longest geometric distance from the negative ideal solution. TOPSIS compares a set of alternatives by normalizing scores for each criterion and calculating the geometric distance between each alternative and the ideal alternative, which is the best score in each criterion. 
The following steps of TOPSIS are required to decide which of the candidate rules is the \emph{most likely rule}.
\begin{enumerate}[leftmargin=*]
    \item Create an evaluation matrix consisting of $m$ alternatives (the rules) and $n$ criteria (the influencing factors), with the intersection of each alternative and criteria given as $x_{ij}$, we therefore have a matrix $(x_{ij})_{m\times n}$.
    \item The matrix values $(x_{ij})_{m\times n}$ are normalized for each column using the following normalization method
    $$r_{\mathit{ij}}=\frac{x_{ij}}{\sqrt{\sum_{k=1}^{m} {x_{kj}^2}}}, i=1,2,...,m, j=1,2,...,n$$
    resulting in the new normalized matrix $R=(r_{ij})_{m \times n}$.
    \item For each column j, we determine the value of the best $A_b$ and worst alternative $A_w$:
        $$A_b=\{\max(t_{ij}|i=1,2,...,m\}\equiv\{t_{\mathit{bj}}|j=1,2,...,n\}$$
        $$A_w=\{\min(t_{ij}|i=1,2,...,m\}\equiv\{t_{\mathit{wj}}|j=1,2,...,n\}$$
    \item Calculate the $L^2$-distance between the target alternative $i$ and the worst condition $A_{w}$
    $$d_{\mathit{ib}}=\sqrt{\sum_{j=1}^{n}(t_{ij}-t_{bj})^2}, i=1,2,...,m$$
    and the distance between the alternative $i$ and the best condition $A_b$
    $$d_{\mathit{iw}}=\sqrt{\sum_{j=1}^{n}(t_{ij}-t_{wj})^2}, i=1,2,...,m$$
    where $d_{\mathit{iw}}$ and $d_{{ib}}$ are $L^2$-norm distances from the target alternative $i$ to the worst and best conditions, respectively.
    \item Calculate the similarity to the worst condition: 
    $$s_{\mathit{iw}}=\frac{d_{\mathit{iw}}}{d_{\mathit{iw}}+d_{\mathit{ib}}}, 0<=s_{\mathit{iw}}<=1, i=1,2,...,m$$
    If alternative $i$ is the worst alternative, $s_{\mathit{iw}}=0$, if it is the best alternative, $s_{\mathit{iw}}=1$.
    \item The alternative with the highest $s_{\mathit{iw}}$ is the \emph{most likely rule} and therefore the one which we will choose to be the \emph{expected rule}.
\end{enumerate}

\subsection{Explanation Generation}
\label{subsection:Framework/pattern}
After the above-mentioned steps, we have determined the fact with its corresponding rule and the foil with its corresponding rule.
To construct a contrastive explanation, we use a set of patterns that embed the fact and the foil into a comprehensive explanation.  
We use a different pattern for each confusing case (CC) (see Table~\ref{table:explanationpatterns}). 

\begin{itemize}[leftmargin=*]
    \item If the \emph{happened event} is a rule (\emph{CC1}), we follow a pattern that explains the contrast between the actions of the \emph{happend rule} ($A_{\mathit{HR}}$) and the actions of the \emph{expected rule} ($A_{\mathit{ER}}$). It also provides a reason why the \emph{happened rule} was fired as it is aware of the preconditions of the \emph{happened rule} ($P_{\mathit{HR}}$).
    \item If no event happened (\emph{CC2}), we follow a pattern, which states that the actions of the \emph{expected rule} ($A_{\mathit{ER}}$) did not occur because the preconditions of this rule ($P_{\mathit{ER}}$) were not met.
    \item If the \emph{happened event} is an error (\emph{CC3}), we follow a pattern that states that the expected actions ($A_{\mathit{ER}}$) did not happen because of the error ($\mathit{HE}$).
\end{itemize}

To streamline the grammatical structure of the final explanation, we provide an LLM with the pattern and the respective constituents and prompt it to generate a grammatically correct explanation without changing its meaning.

\begin{table}
  \caption{Explanation Patterns by \emph{Confusing Case}}
  \label{table:explanationpatterns}
  \centering
  \begin{tabular}{@{}ll@{}}
    \toprule
    \textbf{Case} & \textbf{Pattern}\\
    \midrule
    CC1 & $<A_{HR}>$ occurred instead of $<A_{ER}>$ because $<P_{HR}>$\\
    CC2 & $<A_{ER}>$ did not occur because $<\neg P_{ER}>$\\
    CC3 & $<A_{ER}>$ did not occur because error $<HE>$ occurred.\\
    \bottomrule
  \end{tabular}
\end{table}

\section{Implementation and First Evaluation}

We have implemented the approach as a plugin\footnote{\url{https://github.com/ExmartLab/SmartEx-Engine/tree/contrastive}} for our \emph{SmartEx} explanation engine. \emph{SmartEx} is an extensible framework for adding explainability to smart environment systems. It integrates with Home Assistant\footnote{\url{https://www.home-assistant.io/}}, a popular rule-based engine for home automation, and provides explanation detection and generation capabilities. 
Our prototype extends \emph{SmartEx} by adding explanations with a contrastive element.

To test the feasibility of our approach, we have created a set of test cases for our prototype. 
This prototype has undergone extensive testing in our \emph{smart environments lab}. The lab uses Home Assistant to automate tasks in its smart environment. It contains a context manager, a service outside of Home Assistant and \emph{SmartEx}, that provides contextual information about users.

We utilize two methods of how users can request explanations in the \emph{smart explanation lab}: Scanning an NFC tag attached to a device or filling out a form for explanation requests available as a widget on one of Home Assistant's dashboards. With both methods, the user gives the device they are requesting the explanation for as input at the beginning of the generation process. When this generation is completed, the explanation can be presented to users in several forms: It can be read out loud by the lab's voice controller and can be displayed on the dashboard of Home Assistant on mobile applications, or in web browsers. 

\subsection{Evaluation Scenario}
The underlying scenario for our test cases is a typical office situation. At the door to the office meeting room, an LED status light is installed, which is connected to the smart office management system. Several rules have actions that change the state of the status light. Table~\ref{table:rules} provides an overview of the rules that are important for the scenario. Alice and Bob are coworkers in the office and can set rules for the smart office manager. Additionally, to demonstrate a scenario with \emph{CC3}, an \emph{incorrect installation error} is also part of the test cases. 

\begin{table}
  \caption{Rules of the test scenario}
  \label{table:rules}
  \centering
  \setlength{\extrarowheight}{3pt}
  \begin{tabularx}{\columnwidth}{@{}p{2cm}Xl@{}}
    \toprule
    \textbf{Rule} & \textbf{Description} & \textbf{Owner}\\
    \midrule
    Meeting room not occupied & If the meeting room is not occupied, the status light is turned green & Alice\\
    Meeting room occupied & If the meeting room is occupied, the status light is turned orange & Alice\\
    Rain at lunch & If rain is expected at lunchtime, all status lights are turned blue & Bob\\
    Sun at lunch & If sunny weather is expected at lunchtime, all status lights are turned orange & Bob\\
    Danger & If a dangerous situation arises, all status lights are turned orange & Alice\\
    Closing time & In the evening, all lights are turned off & Alice\\
    \bottomrule
  \end{tabularx}
\end{table}

\subsection{Test Cases}
The test cases listed in Table~\ref{table:testcases} show real-world examples of how our prototype generates contrastive explanations. They highlight the impact of the context on the determination of the expected rule and the generation of the explanation. The test cases show that the prototype delivers context-aware contrastive explanations for a variety of scenarios. Each test case highlights a different aspect that the framework considers when generating the explanation. 

\begin{table}
  \caption{Test cases}
  \label{table:testcases}
  \centering
  \begin{tabularx}{\columnwidth}{@{}p{0.5cm}lXX@{}}
    \toprule
    \textbf{Test} & \textbf{Explainee} & \textbf{Fact} & \textbf{Foil}\\
    \midrule
    1 & Alice & Orange light (meeting room occupied) & Green light (meeting room not occupied)\\
    2 & Bob & Meeting Room occupied & Rain at lunch\\
    3 & Bob & Error & Meeting Room occupied\\
    4 & Alice & Nothing & Meeting Room not occupied\\
    \bottomrule
  \end{tabularx}
\end{table}

\textbf{Test Case 1.}
\label{paragraph:testcase1}
In test case 1, a meeting is held in the meeting room. Most of the people leave the room but a few stay in the room. The \emph{Meeting Room occupied} rule is still in effect and keeps the status light orange. Alice sees the group of people in the hallway after exiting the meeting room. She takes this as a sign that the meeting is finished and wonders why the status light is still orange. According to Table~\ref{table:testcases}, we assume that Alice expects the light to be green because she expected the rule \emph{Meeting Room not occupied} to fire. After she asks \emph{SmartEx} for an explanation, the system first determines the \emph{fact} and finds that the rule \emph{Meeting Room occupied} is the most recent event that happened regarding the status light. As the \emph{happened event} is a rule (\emph{CC1}), the system uses all four expectation factors to determine the \emph{expected rule}. The rules \emph{meeting room occupied}, \emph{sun at lunch} and \emph{danger} are ruled out as candidates as they are either the \emph{happened rule} or they have the same action as the \emph{happened rule}. Each remaining rule is given a score from every expectation factor.  Table~\ref{table:decisionmatrix1} summarizes the scores of all expectation factors. The scores for frequency and explanation occurrence are just random assumptions.

\begin{table}
  \caption{Expected rule scores for test case 1}
  \label{table:decisionmatrix1}
  \centering
  \begin{tabular}{@{}p{3.5cm}rrrr@{}}
    \toprule
    \textbf{Rule} & \textbf{Prec.\ Sim.} & \textbf{Own.} & \textbf{Freq.} & \textbf{Occ.}\\
    \midrule
    Meeting room not occupied & 0.333 & 1 & 65 & 3\\
    Rain at lunch & 0 & 0 & 4 & 4\\
    Closing time & 0 & 1 & 90 & 0\\
    \bottomrule
  \end{tabular}
\end{table}

TOPSIS evaluates this decision matrix with the steps outlined in Section~\ref{subsec:Framework/rating}. In Table~\ref{table:topsisresults1}, the distances of the alternatives to the best alternative ($d_{ib}$) and worst alternative ($d_{\mathit{iw}}$) and the final performance score $s_{\mathit{iw}}$ are shown for the candidate rules (rounded to 3 decimal places). Because the performance score for the \emph{Meeting room not occupied} rule is the highest, it is the \emph{most likely rule} and the system therefore determines it to be the \emph{expected rule}.

\begin{table}
  \caption{TOPSIS results for test case 1}
  \label{table:topsisresults1}
  \centering
  \begin{tabular}{lrrr}
    \toprule
    \textbf{Rule} & \textbf{$d_{ib}$} & \textbf{$d_{\mathit{iw}}$} & \textbf{$s_{\mathit{iw}}$}\\
    \midrule
    Meeting room not occupied & 0.075 & 0.368 & \textbf{0.830}\\
    Rain at lunch & 0.362 & 0.2 & 0.356\\
    Closing time & 0.320 & 0.262 & 0.450\\
    \bottomrule
  \end{tabular}
\end{table}

Thus, the system correctly determines that \emph{Meeting room not occupied} is the \emph{expected rule} as it is the one Alice expected originally. The system then generates the following contrastive explanation according to the pattern for CC1: \textit{For the meeting room status light, an orange hue occurred instead of the expected green light because the occupancy of the meeting room was detected, and the contact sensor for the meeting room door is turned off.}

\textbf{Test Case 2.}
Bob set a rule that turns all status lights in the office to blue if rain is expected at lunchtime to remind everyone to take a coat or umbrella when going out for lunch. Now it is lunchtime and he notices that it is raining outside. A meeting is currently going on in the meeting room and Alice's rule (Meeting room occupied) turned the meeting room status light orange. As Bob walks past the meeting room, he wonders and asks \emph{SmartEx} for an explanation as he is confused by the orange status light. The system first determines the \emph{fact} and finds that the rule \emph{Meeting Room occupied} is the event that happened regarding the meeting room status light. As the \emph{happened event} is a rule (\emph{CC1}), the system uses all four expectation measures to determine the \emph{most likely rule}. 
The rest of the computation is similar to test case 1. In the end, the \emph{rain at lunch} rule is determined to be the expected rule and the the following explanation is generated: 
\textit{For the meeting room status light, an orange hue occurred instead of the expected blue light because the occupancy of the meeting room was detected, and the contact sensor for the meeting room door is off.}

\textbf{Test Case 3.}
After some maintenance work in the office, the status light was not installed properly again and therefore is not connected correctly. Bob is late to a meeting in the meeting room. As he approaches the meeting room, he wonders why the light is off and not orange because he knows the meeting already started. He asks \emph{SmartEx} for an explanation about the light. The system finds that the \emph{happened event} is an \emph{incorrect installation error}. As an error happened \emph{CC3}, the system looks at the Home Assistant log and determines that the \emph{expected rule} is the last rule that was fired with an action involving the status light. At the start of the meeting, the system fired the \emph{Meeting room occupied} rule and tried to turn the status light orange but as it could not be reached because of the faulty connection, the error occurred. Knowing the \emph{fact} and the \emph{foil}, \emph{SmartEx} generates the explanation based on the pattern for CC3: \textit{The orange light did not occur because an incorrect installation error occurred in the meeting room status light.}

\textbf{Test Case 4.}
One day, Alice comes into the office in the morning and wonders why the status light is off. She expects it to be green as she programmed a rule (meeting room not occupied) to show if the room is available. What she does not know is that the door was not fully closed so the contact sensor assumed the door was still open. She asks \emph{SmartEx} for an explanation. The system finds that no rule has been fired that would have impacted the light (\emph{CC2}), so it uses the expectation measures to determine the \emph{expected rule}. Precondition similarity cannot be assessed as no rule was fired to compare the preconditions of the candidate rules.

Again, in this case, the system correctly determines that \emph{Meeting room not occupied} is the \emph{expected rule} as it is the one Alice expected originally. The system generates an explanation based on the pattern for CC2: \textit{For the meeting room status light, the green light did not occur because there was motion in the meeting room or the contact sensor for the meeting room was not closed.}

\vspace{-0.5em}
\subsection{Limitations}
\label{subsec:Framework/Discussion}
While we cover the entire process of the generation of the explanation, one area that is not addressed is the detection of an explanation need. The process of the framework starts with the existence of a need for an explanation. This can either be voiced by the explainee personally or be detected via a separate system.

The occasion that a rule does not exist anymore in the system is not covered by this framework. It would be a subcase of \emph{CC2} as a non-existing rule cannot be fired and could be part of future work. The explainable system could have a database of past rules and merge those with the current rules into the set of candidate rules for the determination of the \emph{expected rule} in \emph{CC2}.

In general, TOPSIS also allows weighting criteria differently. For simplicity, we decided to weigh all criteria equally. For future work, the weighting schemes proposed and analyzed by \cite{olson2004comparison} could assist in choosing different weights for the TOPSIS criteria.

Overall, this framework is a novel approach for the generation of context-aware contrastive explanations in smart environments. It delivers a comprehensive process from the determination of the \emph{fact} and \emph{foil} to the construction of the actual explanation as a sentence.

\vspace{-0.5em}
\section{Related Work}

Currently, there only exist a few approaches for explaining the behavior of cyber-physical systems (CPS). 
Drechsler~et~al.~\cite{drechsler2018} sketch the first steps toward a conceptual framework for self-explaining CPS. They propose to add a layer for self-explanation, which includes an abstract model of the system. They also propose adjusting the granularity of explanations for different predefined user groups. The generated explanations are cause-effect chains for observable actions using the abstract model. Users can access these chains to understand the cause of actions. 

Chiyah Garcia et al.~\cite{ChiyahHRI18} present a modified version of fault trees that can be used to extract explanations. They call this the \textit{model of autonomy}, which captures possible states of a system and the potential reasons for this state. This model has to be developed manually by experts.
In contrast, Plambeck et al.~\cite{plambeck22} propose to learn dependencies on external influences for CPS with decision trees. This approach automatically identifies relevant influences and extracts data-related explanations of the system behavior. Such decision trees were already proven to be valuable for generating \textit{why} and \textit{why-not} explanations~\cite{lim2009and}.

Houz{\'e} et al.~\cite{houze2022generic} propose a modular architecture for self\-/explainable smart homes. The architecture consists of local explanatory components (LECs) that monitor single components and explain their behavior in terms of component-independent abstractions – predicates and events. A generic central component, called \textit{Spotlight}, coordinates the LECs to generate system-wide explanations.

Agrawal and Cleland-Huang~\cite{Agrawal2021} examine the User Interface (UI) design trade-offs associated with providing timely and detailed explanations of autonomous behavior for swarms of small Unmanned Aerial Systems (sUAS). They analyze the impact of UI design choices on human awareness of the situation. They provide actionable guidelines for effectively explaining the autonomous behavior of multiple sUAS.

\vspace{-0.5em}
\section{Conclusions}

We proposed an approach for generating contrastive explanations in rule-based systems. It is designed to generate an explanation based on the current situation of the user. The approach considers several contextual factors to determine the \emph{foil}, the situation that the user expected instead of the \emph{fact}. The approach identifies three basic confusing cases (\emph{CC}) that can lead to a contrastive situation where the user expected a different outcome. 
The factors used by the approach include precondition similarity, ownership, frequency, and occurrence. The scores assigned to the candidate rules by these measures are being evaluated by a rating system to arrive at a final decision about the \emph{most likely rule}. We used TOPSIS as a multi-criteria decision analysis method particularly suited for this type of problem.
Overall, this framework provides a comprehensive and intelligent approach to generating context-aware contrastive explanations in smart environments.
The approach does not address the detection of an explanation need, which can either be voiced by the explainee personally or detected via a separate system.

\bibliographystyle{ACM-Reference-Format}
\bibliography{references}


\begin{thebibliography}{31}


\ifx \showCODEN    \undefined \def \showCODEN     #1{\unskip}     \fi
\ifx \showDOI      \undefined \def \showDOI       #1{#1}\fi
\ifx \showISBNx    \undefined \def \showISBNx     #1{\unskip}     \fi
\ifx \showISBNxiii \undefined \def \showISBNxiii  #1{\unskip}     \fi
\ifx \showISSN     \undefined \def \showISSN      #1{\unskip}     \fi
\ifx \showLCCN     \undefined \def \showLCCN      #1{\unskip}     \fi
\ifx \shownote     \undefined \def \shownote      #1{#1}          \fi
\ifx \showarticletitle \undefined \def \showarticletitle #1{#1}   \fi
\ifx \showURL      \undefined \def \showURL       {\relax}        \fi
\providecommand\bibfield[2]{#2}
\providecommand\bibinfo[2]{#2}
\providecommand\natexlab[1]{#1}
\providecommand\showeprint[2][]{arXiv:#2}

\bibitem[Agrawal and Cleland-Huang(2021)]%
        {Agrawal2021}
\bibfield{author}{\bibinfo{person}{Ankit Agrawal} {and} \bibinfo{person}{Jane
  Cleland-Huang}.} \bibinfo{year}{2021}\natexlab{}.
\newblock \showarticletitle{Explaining Autonomous Decisions in Swarms of
  Human-on-the-Loop Small Unmanned Aerial Systems}.
\newblock \bibinfo{journal}{\emph{{AAAI} Conference on Human Computation and
  Crowdsourcing}}  \bibinfo{volume}{9} (\bibinfo{year}{2021}),
  \bibinfo{pages}{15--26}.
\newblock
\urldef\tempurl%
\url{https://doi.org/10.1609/hcomp.v9i1.18936}
\showDOI{\tempurl}


\bibitem[Alam et~al\mbox{.}(2012)]%
        {alam2012review}
\bibfield{author}{\bibinfo{person}{Muhammad~Raisul Alam},
  \bibinfo{person}{Mamun Bin~Ibne Reaz}, {and} \bibinfo{person}{Mohd
  Alauddin~Mohd Ali}.} \bibinfo{year}{2012}\natexlab{}.
\newblock \showarticletitle{A review of smart homes—Past, present, and
  future}.
\newblock \bibinfo{journal}{\emph{IEEE Transactions on Systems, Man, and
  Cybernetics, part C (applications and reviews)}} \bibinfo{volume}{42},
  \bibinfo{number}{6} (\bibinfo{year}{2012}), \bibinfo{pages}{1190–--1203}.
\newblock
\urldef\tempurl%
\url{https://doi.org/10.1109/tsmcc.2012.2189204}
\showDOI{\tempurl}


\bibitem[Artelt et~al\mbox{.}(2021)]%
        {artelt2021contrastive}
\bibfield{author}{\bibinfo{person}{Andr{\'e} Artelt}, \bibinfo{person}{Fabian
  Hinder}, \bibinfo{person}{Valerie Vaquet}, \bibinfo{person}{Robert Feldhans},
  {and} \bibinfo{person}{Barbara Hammer}.} \bibinfo{year}{2021}\natexlab{}.
\newblock \bibinfo{booktitle}{\emph{Contrastive Explanations for Explaining
  Model Adaptations}}.
\newblock \bibinfo{publisher}{Springer International Publishing},
  \bibinfo{pages}{101--–112}.
\newblock
\showISBNx{9783030850302}
\showISSN{1611-3349}
\urldef\tempurl%
\url{https://doi.org/10.1007/978-3-030-85030-2_9}
\showDOI{\tempurl}


\bibitem[Blumreiter et~al\mbox{.}(2019)]%
        {Blumreiter2019}
\bibfield{author}{\bibinfo{person}{Mathias Blumreiter}, \bibinfo{person}{Joel
  Greenyer}, \bibinfo{person}{Francisco~Javier Chiyah~Garcia},
  \bibinfo{person}{Verena Klos}, \bibinfo{person}{Maike Schwammberger},
  \bibinfo{person}{Christoph Sommer}, \bibinfo{person}{Andreas Vogelsang},
  {and} \bibinfo{person}{Andreas Wortmann}.} \bibinfo{year}{2019}\natexlab{}.
\newblock \showarticletitle{Towards Self-Explainable Cyber-Physical Systems}.
  In \bibinfo{booktitle}{\emph{ACM/IEEE 22nd International Conference on Model
  Driven Engineering Languages and Systems Companion (MODELS-C)}}.
  \bibinfo{publisher}{IEEE}.
\newblock
\urldef\tempurl%
\url{https://doi.org/10.1109/models-c.2019.00084}
\showDOI{\tempurl}


\bibitem[Brunotte et~al\mbox{.}(2022)]%
        {Brunotte2022}
\bibfield{author}{\bibinfo{person}{Wasja Brunotte}, \bibinfo{person}{Larissa
  Chazette}, \bibinfo{person}{Verena Kl\"{o}s}, {and} \bibinfo{person}{Timo
  Speith}.} \bibinfo{year}{2022}\natexlab{}.
\newblock \bibinfo{booktitle}{\emph{Quo Vadis, Explainability? – A Research
  Roadmap for Explainability Engineering}}.
\newblock \bibinfo{publisher}{Springer International Publishing},
  \bibinfo{pages}{26--–32}.
\newblock
\showISBNx{9783030984649}
\showISSN{1611-3349}
\urldef\tempurl%
\url{https://doi.org/10.1007/978-3-030-98464-9_3}
\showDOI{\tempurl}


\bibitem[Chazette and Schneider(2020)]%
        {chazette2020explainability}
\bibfield{author}{\bibinfo{person}{Larissa Chazette} {and}
  \bibinfo{person}{Kurt Schneider}.} \bibinfo{year}{2020}\natexlab{}.
\newblock \showarticletitle{Explainability as a non-functional requirement:
  challenges and recommendations}.
\newblock \bibinfo{journal}{\emph{Requirements Engineering}}
  \bibinfo{volume}{25}, \bibinfo{number}{4} (\bibinfo{year}{2020}),
  \bibinfo{pages}{493--–514}.
\newblock
\showISSN{1432-010X}
\urldef\tempurl%
\url{https://doi.org/10.1007/s00766-020-00333-1}
\showDOI{\tempurl}


\bibitem[Chen et~al\mbox{.}(2021)]%
        {chen2021kace}
\bibfield{author}{\bibinfo{person}{Qianglong Chen}, \bibinfo{person}{Feng Ji},
  \bibinfo{person}{Xiangji Zeng}, \bibinfo{person}{Feng-Lin Li},
  \bibinfo{person}{Ji Zhang}, \bibinfo{person}{Haiqing Chen}, {and}
  \bibinfo{person}{Yin Zhang}.} \bibinfo{year}{2021}\natexlab{}.
\newblock \showarticletitle{{KACE}: Generating Knowledge Aware Contrastive
  Explanations for Natural Language Inference}. In
  \bibinfo{booktitle}{\emph{59th Annual Meeting of the Association for
  Computational Linguistics and the 11th International Joint Conference on
  Natural Language Processing (Volume 1: Long Papers)}}.
  \bibinfo{publisher}{Association for Computational Linguistics}.
\newblock
\urldef\tempurl%
\url{https://doi.org/10.18653/v1/2021.acl-long.196}
\showDOI{\tempurl}


\bibitem[Dai et~al\mbox{.}(2023)]%
        {dai2023effect}
\bibfield{author}{\bibinfo{person}{Jiaxin Dai}, \bibinfo{person}{Chao Zhang},
  \bibinfo{person}{Dzmitry Aliakseyeu}, \bibinfo{person}{Samantha Peeters},
  {and} \bibinfo{person}{Wijnand~A Ijsselsteijn}.}
  \bibinfo{year}{2023}\natexlab{}.
\newblock \showarticletitle{The Effect of Explanation Design on User Perception
  of Smart Home Lighting Systems: A Mixed-method Investigation}. In
  \bibinfo{booktitle}{\emph{CHI Conference on Human Factors in Computing
  Systems}}. \bibinfo{publisher}{ACM}.
\newblock
\urldef\tempurl%
\url{https://doi.org/10.1145/3544548.3581263}
\showDOI{\tempurl}


\bibitem[Drechsler et~al\mbox{.}(2018)]%
        {drechsler2018}
\bibfield{author}{\bibinfo{person}{Rolf Drechsler}, \bibinfo{person}{Christoph
  Luth}, \bibinfo{person}{Goerschwin Fey}, {and} \bibinfo{person}{Tim
  Guneysu}.} \bibinfo{year}{2018}\natexlab{}.
\newblock \showarticletitle{Towards Self-Explaining Digital Systems: A Design
  Methodology for the Next Generation}. In \bibinfo{booktitle}{\emph{IEEE 3rd
  International Verification and Security Workshop (IVSW)}}.
  \bibinfo{publisher}{IEEE}.
\newblock
\urldef\tempurl%
\url{https://doi.org/10.1109/ivsw.2018.8494900}
\showDOI{\tempurl}


\bibitem[El-Din et~al\mbox{.}(2020)]%
        {el2021smart}
\bibfield{author}{\bibinfo{person}{Doaa~Mohey El-Din},
  \bibinfo{person}{Aboul~Ella Hassanein}, {and} \bibinfo{person}{Ehab~E.
  Hassanien}.} \bibinfo{year}{2020}\natexlab{}.
\newblock \bibinfo{booktitle}{\emph{Smart Environments Concepts, Applications,
  and Challenges}}.
\newblock \bibinfo{publisher}{Springer International Publishing},
  \bibinfo{pages}{493--–519}.
\newblock
\showISBNx{9783030593384}
\showISSN{2197-6511}
\urldef\tempurl%
\url{https://doi.org/10.1007/978-3-030-59338-4_24}
\showDOI{\tempurl}


\bibitem[Ferreira and Monteiro(2020)]%
        {ferreira2020people}
\bibfield{author}{\bibinfo{person}{Juliana~J. Ferreira} {and}
  \bibinfo{person}{Mateus~S. Monteiro}.} \bibinfo{year}{2020}\natexlab{}.
\newblock \bibinfo{booktitle}{\emph{What Are People Doing About {XAI} User
  Experience? A Survey on AI Explainability Research and Practice}}.
\newblock \bibinfo{publisher}{Springer International Publishing},
  \bibinfo{pages}{56–--73}.
\newblock
\showISBNx{9783030497606}
\showISSN{1611-3349}
\urldef\tempurl%
\url{https://doi.org/10.1007/978-3-030-49760-6_4}
\showDOI{\tempurl}


\bibitem[Garcia et~al\mbox{.}(2018)]%
        {ChiyahHRI18}
\bibfield{author}{\bibinfo{person}{Francisco J.~Chiyah Garcia},
  \bibinfo{person}{David~A. Robb}, \bibinfo{person}{X. Liu},
  \bibinfo{person}{Atanas Laskov}, \bibinfo{person}{Patron Patron}, {and}
  \bibinfo{person}{Helen Hastie}.} \bibinfo{year}{2018}\natexlab{}.
\newblock \showarticletitle{Explain Yourself: {A} Natural Language Interface
  for Scrutable Autonomous Robots}. In \bibinfo{booktitle}{\emph{Explainable
  Robotic Systems Workshop (HRI)}}.
\newblock
\urldef\tempurl%
\url{https://doi.org/10.48550/arXiv.1803.02088}
\showDOI{\tempurl}


\bibitem[Houz{\'e} et~al\mbox{.}(2022)]%
        {houze2022generic}
\bibfield{author}{\bibinfo{person}{Etienne Houz{\'e}}, \bibinfo{person}{Ada
  Diaconescu}, \bibinfo{person}{Jean-Louis Dessalles}, {and}
  \bibinfo{person}{David Menga}.} \bibinfo{year}{2022}\natexlab{}.
\newblock \showarticletitle{A generic and modular reference architecture for
  self-explainable smart homes}. In \bibinfo{booktitle}{\emph{IEEE
  International Conference on Autonomic Computing and Self-Organizing Systems
  (ACSOS)}}. IEEE.
\newblock


\bibitem[Hwang et~al\mbox{.}(1993)]%
        {hwang1993new}
\bibfield{author}{\bibinfo{person}{Ching-Lai Hwang}, \bibinfo{person}{Young-Jou
  Lai}, {and} \bibinfo{person}{Ting-Yun Liu}.} \bibinfo{year}{1993}\natexlab{}.
\newblock \showarticletitle{A new approach for multiple objective decision
  making}.
\newblock \bibinfo{journal}{\emph{Computers \& operations research}}
  \bibinfo{volume}{20}, \bibinfo{number}{8} (\bibinfo{year}{1993}).
\newblock


\bibitem[Jacovi et~al\mbox{.}(2021)]%
        {jacovi2021contrastive}
\bibfield{author}{\bibinfo{person}{Alon Jacovi}, \bibinfo{person}{Swabha
  Swayamdipta}, \bibinfo{person}{Shauli Ravfogel}, \bibinfo{person}{Yanai
  Elazar}, \bibinfo{person}{Yejin Choi}, {and} \bibinfo{person}{Yoav
  Goldberg}.} \bibinfo{year}{2021}\natexlab{}.
\newblock \showarticletitle{Contrastive explanations for model
  interpretability}.
\newblock \bibinfo{journal}{\emph{arXiv preprint arXiv:2103.01378}}
  (\bibinfo{year}{2021}).
\newblock


\bibitem[Lewis(1986)]%
        {lewis1986causal}
\bibfield{author}{\bibinfo{person}{David Lewis}.}
  \bibinfo{year}{1986}\natexlab{}.
\newblock \showarticletitle{Causal Explanation}.
\newblock In \bibinfo{booktitle}{\emph{Philosophical Papers Vol. Ii}},
  \bibfield{editor}{\bibinfo{person}{David Lewis}} (Ed.).
  \bibinfo{publisher}{Oxford University Press}, \bibinfo{pages}{214--240}.
\newblock


\bibitem[Liao et~al\mbox{.}(2020)]%
        {liao2020questioning}
\bibfield{author}{\bibinfo{person}{{Q. Vera} Liao}, \bibinfo{person}{Daniel
  Gruen}, {and} \bibinfo{person}{Sarah Miller}.}
  \bibinfo{year}{2020}\natexlab{}.
\newblock \showarticletitle{Questioning the AI: informing design practices for
  explainable AI user experiences}. In \bibinfo{booktitle}{\emph{CHI conference
  on human factors in computing systems}}.
\newblock


\bibitem[Lim et~al\mbox{.}(2009)]%
        {lim2009and}
\bibfield{author}{\bibinfo{person}{Brian~Y Lim}, \bibinfo{person}{Anind~K Dey},
  {and} \bibinfo{person}{Daniel Avrahami}.} \bibinfo{year}{2009}\natexlab{}.
\newblock \showarticletitle{Why and why not explanations improve the
  intelligibility of context-aware intelligent systems}. In
  \bibinfo{booktitle}{\emph{SIGCHI conference on human factors in computing
  systems}}.
\newblock


\bibitem[Lipton(1990)]%
        {lipton1990contrastive}
\bibfield{author}{\bibinfo{person}{Peter Lipton}.}
  \bibinfo{year}{1990}\natexlab{}.
\newblock \showarticletitle{Contrastive explanation}.
\newblock \bibinfo{journal}{\emph{Royal Institute of Philosophy Supplements}}
  \bibinfo{volume}{27} (\bibinfo{year}{1990}).
\newblock


\bibitem[Miller(2019)]%
        {miller2019explanation}
\bibfield{author}{\bibinfo{person}{Tim Miller}.}
  \bibinfo{year}{2019}\natexlab{}.
\newblock \showarticletitle{Explanation in artificial intelligence: Insights
  from the social sciences}.
\newblock \bibinfo{journal}{\emph{Artificial intelligence}}
  \bibinfo{volume}{267} (\bibinfo{year}{2019}).
\newblock


\bibitem[Miller(2021)]%
        {miller2021contrastive}
\bibfield{author}{\bibinfo{person}{Tim Miller}.}
  \bibinfo{year}{2021}\natexlab{}.
\newblock \showarticletitle{Contrastive explanation: A structural-model
  approach}.
\newblock \bibinfo{journal}{\emph{The Knowledge Engineering Review}}
  \bibinfo{volume}{36} (\bibinfo{year}{2021}).
\newblock


\bibitem[Nandi and Ernst(2016)]%
        {nandi2016automatic}
\bibfield{author}{\bibinfo{person}{Chandrakana Nandi} {and}
  \bibinfo{person}{Michael~D Ernst}.} \bibinfo{year}{2016}\natexlab{}.
\newblock \showarticletitle{Automatic trigger generation for rule-based smart
  homes}. In \bibinfo{booktitle}{\emph{ACM Workshop on Programming Languages
  and Analysis for Security}}.
\newblock


\bibitem[Olson(2004)]%
        {olson2004comparison}
\bibfield{author}{\bibinfo{person}{David~L Olson}.}
  \bibinfo{year}{2004}\natexlab{}.
\newblock \showarticletitle{Comparison of weights in TOPSIS models}.
\newblock \bibinfo{journal}{\emph{Mathematical and Computer Modelling}}
  \bibinfo{volume}{40}, \bibinfo{number}{7} (\bibinfo{year}{2004}).
\newblock


\bibitem[Plambeck et~al\mbox{.}(2022)]%
        {plambeck22}
\bibfield{author}{\bibinfo{person}{Swantje Plambeck},
  \bibinfo{person}{Gorschwin Fey}, \bibinfo{person}{Jakob Schyga},
  \bibinfo{person}{Johannes Hinckeldeyn}, {and} \bibinfo{person}{Jochen
  Kreutzfeldt}.} \bibinfo{year}{2022}\natexlab{}.
\newblock \showarticletitle{Explaining Cyber-Physical Systems Using Decision
  Trees}. In \bibinfo{booktitle}{\emph{2nd International Workshop on
  Computation-Aware Algorithmic Design for Cyber-Physical Systems
  ({CAADCPS})}}. \bibinfo{publisher}{{IEEE}}.
\newblock
\urldef\tempurl%
\url{https://doi.org/10.1109/caadcps56132.2022.00006}
\showDOI{\tempurl}


\bibitem[Reisinger et~al\mbox{.}(2022)]%
        {reisinger2022user}
\bibfield{author}{\bibinfo{person}{Michaela~R Reisinger},
  \bibinfo{person}{Sebastian Prost}, \bibinfo{person}{Johann Schrammel}, {and}
  \bibinfo{person}{Peter Fr{\"o}hlich}.} \bibinfo{year}{2022}\natexlab{}.
\newblock \showarticletitle{User requirements for the design of smart homes:
  dimensions and goals}.
\newblock \bibinfo{journal}{\emph{Journal of Ambient Intelligence and Humanized
  Computing}} (\bibinfo{year}{2022}).
\newblock


\bibitem[Roese(1997)]%
        {roese1997counterfactual}
\bibfield{author}{\bibinfo{person}{Neal~J Roese}.}
  \bibinfo{year}{1997}\natexlab{}.
\newblock \showarticletitle{Counterfactual thinking.}
\newblock \bibinfo{journal}{\emph{Psychological bulletin}}
  \bibinfo{volume}{121}, \bibinfo{number}{1} (\bibinfo{year}{1997}).
\newblock


\bibitem[Sadeghi et~al\mbox{.}(2024)]%
        {sadeghi2024}
\bibfield{author}{\bibinfo{person}{Mersedeh Sadeghi}, \bibinfo{person}{Lars
  Herbold}, \bibinfo{person}{Max Unterbusch}, {and} \bibinfo{person}{Andreas
  Vogelsang}.} \bibinfo{year}{2024}\natexlab{}.
\newblock \showarticletitle{{SmartEx}: A Framework for Generating User-Centric
  Explanations in Smart Environments}. In \bibinfo{booktitle}{\emph{IEEE
  International Conference on Pervasive Computing and Communications
  (PerCom)}}. \bibinfo{publisher}{IEEE}.
\newblock


\bibitem[Sadeghi et~al\mbox{.}(2021)]%
        {Sadeghi21}
\bibfield{author}{\bibinfo{person}{Mersedeh Sadeghi}, \bibinfo{person}{Verena
  Klös}, {and} \bibinfo{person}{Andreas Vogelsang}.}
  \bibinfo{year}{2021}\natexlab{}.
\newblock \showarticletitle{Cases for Explainable Software Systems:
  Characteristics and Examples}. In \bibinfo{booktitle}{\emph{IEEE 29th
  International Requirements Engineering Conference Workshops (REW)}}.
\newblock
\urldef\tempurl%
\url{https://doi.org/10.1109/REW53955.2021.00033}
\showDOI{\tempurl}


\bibitem[Stepin et~al\mbox{.}(2021)]%
        {stepin2021survey}
\bibfield{author}{\bibinfo{person}{Ilia Stepin}, \bibinfo{person}{Jose~M
  Alonso}, \bibinfo{person}{Alejandro Catala}, {and}
  \bibinfo{person}{Mart{\'\i}n Pereira-Fari{\~n}a}.}
  \bibinfo{year}{2021}\natexlab{}.
\newblock \showarticletitle{A survey of contrastive and counterfactual
  explanation generation methods for explainable artificial intelligence}.
\newblock \bibinfo{journal}{\emph{IEEE Access}}  \bibinfo{volume}{9}
  (\bibinfo{year}{2021}).
\newblock


\bibitem[Winikoff(2018)]%
        {winikoff2018towards}
\bibfield{author}{\bibinfo{person}{Michael Winikoff}.}
  \bibinfo{year}{2018}\natexlab{}.
\newblock \showarticletitle{Towards trusting autonomous systems}. In
  \bibinfo{booktitle}{\emph{5th International Workshop on Engineering
  Multi-Agent Systems (EMAS 2017)}}. Springer.
\newblock


\bibitem[Zulqarnain et~al\mbox{.}(2020)]%
        {zulqarnain2020application}
\bibfield{author}{\bibinfo{person}{RM Zulqarnain}, \bibinfo{person}{M Saeed},
  \bibinfo{person}{N Ahmad}, \bibinfo{person}{F Dayan}, {and}
  \bibinfo{person}{B Ahmad}.} \bibinfo{year}{2020}\natexlab{}.
\newblock \showarticletitle{Application of TOPSIS method for decision making}.
\newblock \bibinfo{journal}{\emph{IJSRMSS International Journal of Scientific
  Research in Mathematical and Statistical Sciences}} (\bibinfo{year}{2020}).
\newblock


\end{thebibliography}

\end{document}